\documentclass[]{article}
\usepackage{listings}
\usepackage{graphicx}
\usepackage{xcolor}
\usepackage{systeme, mathtools}
\usepackage{supertabular}
\usepackage{hhline}
\usepackage{amsmath}
\usepackage{adjustbox}
\usepackage{amsfonts}
\usepackage{float}
\usepackage{array}
\usepackage{listings}
\usepackage[justification=centering]{caption}
\usepackage[title]{appendix}
\usepackage{courier}
\usepackage{hyperref}
\usepackage{tikz}
\usepackage{csquotes}

\usepgflibrary{arrows}
\usetikzlibrary{shapes,snakes}
\usepackage{pgfplots}
\usepackage{amsmath,amsfonts,amssymb}
\usetikzlibrary{arrows.meta}

\makeatletter
\newcommand*{\centerfloat}{%
	\parindent \z@
	\leftskip \z@ \@plus 1fil \@minus \textwidth
	\rightskip\leftskip
	\parfillskip \z@skip}
\makeatother

\definecolor{codecolors}{RGB}{ 219,225,226}
\definecolor{codecolors_inline}{RGB}{ 230.1000,  234.3000 , 235.0000}

\lstdefinestyle{mycodestyle}{
 backgroundcolor = \color{codecolors},
                   language = Python,
                   xleftmargin = 0.5cm,
                   framexleftmargin = 2em, mathescape=true
}

\definecolor{y_box}{RGB}{ 171,131,174}
\definecolor{g_box}{RGB}{56,168,83}

\definecolor{gray_box}{rgb}{0.85, 0.85, 0.85}

\definecolor{mycolor}{RGB}{202,230,223}
\definecolor{mycolor3}{RGB}{10,103,79}
\definecolor{mycolor2}{RGB}{171,131,174}

\tikzset{
    my_latent/.style={circle, draw=mycolor3, thick, fill=mycolor},
    my_obs/.style={rectangle, draw = mycolor3, thick, fill=mycolor},
    my_manif/.style={},
    my_arrow/.style={-{Latex[length=2mm]}, mycolor3, thick},
    my_arrow_mp/.style={-{Latex[length=1.5mm]}, shorten <=2pt, black},
	my_covariance/.style={{Latex[length=2mm]}-{Latex[length=2mm]}, 
	 mycolor2, thick},
	my_covariance_mp/.style={{Latex[length=1.5mm]}-{Latex[length=1.5mm]}, 
	 mycolor2, thick}
}

\title{\textbf{semopy}:  A Python package for Structural Equation Modeling}
\author{Georgy Meshcheryakov, Anna A. Igolkina}

\begin{document}

\maketitle

\begin{abstract}
	Structural equation modelling (SEM) is a multivariate statistical technique for estimating complex relationships between observed and latent variables. Although numerous SEM packages exist, each of them has limitations. Some packages are not free or open-source; the most popular package not having this disadvantage is \textbf{lavaan}, but it is written in R language, which is behind current mainstream tendencies that make it harder to be incorporated into developmental pipelines (i.e. bioinformatical ones). Thus we developed the Python package \textbf{semopy} to satisfy those criteria. The paper provides detailed examples of package usage and explains it's inner clockworks. Moreover, we developed the unique generator of SEM models to extensively test SEM packages and demonstrated that \textbf{semopy} significantly outperforms \textbf{lavaan} in execution time and accuracy.
\end{abstract}

\section{Introduction}

Structural Equation Modelling (SEM) can be defined as a diverse set of tools and approaches for describing and estimating causal relationships between variables, whether they be observable or latent. The very early beginnings of SEM models (Path analysis) were established in the first half of the 20th century by a geneticist and statistician Sewall Green Wright and deal with only observed variables \cite{wrights.1921}. Over the years, approaches for working with latent variables have been developed (e.g. factor analysis), and in modern SEM models, researchers can specify complex hybrids of path analysis models, confirmatory path analysis models (CFA) \cite{jkcfa.1969} and multivariate regression models.  
Being an umbrella term over the statistical approaches, SEM utilises particular statistical methods to estimate relationships between variables describing a variance-covariance structure of the data via model parameters \cite{bollenk.1989}.

The first SEM model is LISREL (linear structural relations) and contains two parts: (i) the structural part links latent variables to each other via a system of linear equations; (ii) the measurement part specifies linear influences of latent variables to observed variables \cite{bollenk.1989}. Let $\eta$ be a vector of latent variables, $y$ be a vector of observed (manifest) variables, then the two parts of LISREL model are as follows:

\begin{equation} \label{lisrel}
\begin{cases}
\eta = \mathrm{B} \eta + \varepsilon, \\
y = \Lambda \eta + \delta,
\end{cases}
\end{equation}
where $\mathrm{B}$ and $\mathrm{\Lambda}$ are matrices with linear parameters, $\varepsilon$ and $\delta$ are independent error terms. Following the LISREL notation, complex SEM models are traditionally split into two parts: structural and measurement ones. The \textbf{semopy} package supports a general SEM model allowing the presence of observed variables in the structural part, as well as provides support for ordinal variables.

In this paper, we discuss the prerequisites for development of \textbf{semopy} package, explain the user-friendly syntax used for specifying an SEM model and provide quick start information on the usage of the \textbf{semopy} package. Then, we provide implementation details such as an underlying mathematical model, heuristics for choosing starting values, provide a list of objective functions and optimization techniques at user's disposal. We also explain in brief statistics such as p-values and fit indices and introduce a testing framework that generates random sets of models and data. In the end, we compare \textbf{semopy}  to \textbf{lavaan} \cite{lavcit} -- the-state-of-the-art CRAN R package that implements SEM functionality.

%
\subsection{Why do we need semopy?}

Nowadays SEM is widely used in the fields of economics, psychology, sociology and bioinformatics \cite{Igolkina2018, pugesek.2009, grace.2006} and there is a number of software working with SEM models. Most of them are either commercially distributed and non-open-sourced or do not cover the whole set of popular programming languages, e.g. Python. 


For instance, \textbf{LISREL}\cite{lisrel}, \textbf{Mplus}\cite{mplus} and \textbf{EQS} \cite{eqs} are proprietary and commercial softwares, hence, any adjustments to them to satisfy researchers needs are not possible. \textbf{OpenMx}\cite{openmx}, \textbf{sem}\cite{sem}, \textbf{lavaan} are free and open-source popular CRAN packages, but they are all written in R. The only Python package is \textbf{pypsy}\cite{pypsy}, but it is limited to basic SEM functionality and by the lack of documentation. Our reasons for developing the new SEM package, \textbf{semopy}, are as following:

\begin{itemize}
	\item the need for SEM package which could be easily integrated into developmental and research pipelines in Python (especially into bioinformatic ones) \cite{Pypl, Tiobe};
	\item the wish to outperform in execution time and accuracy the most cited open-source package, \textbf{lavaan};
	\item the lack of a profound testing technique for new SEM methods and approaches.
\end{itemize}

\subsection{Model syntax}
To specify SEM models, The \textbf{semopy} uses the \textbf{lavaan} syntax, which is natural to describe regression models in R. The syntax supports three operator symbols characterising relationships between variables:
\begin{itemize}
	\item \lstinline{~} to specify structural part,
	\item \lstinline{=~} to specify measurement part,
	\item \lstinline{~~} to specify common variance between variables.
\end{itemize}

\begin{figure}[H]
	\centering
	\begin{tikzpicture}
	\begin{scope}[scale=1.3, transform shape]
	
	\node[my_manif] (y1) at (1,10) {$y_1$};
	\node[my_manif] (y2) at (1,9) {$y_2$};
	\node[my_manif] (y3) at (1,8) {$y_3$};
	\node[my_manif] (y4) at (4,7) {$y_4$};
	\node[my_manif] (y5) at (4,6) {$y_5$};
	\node[my_manif] (y6) at (5,7) {$y_6$};
	
	\node[my_obs] (x1) at (2,7) {$x_1$};
	\node[my_obs] (x2) at (2,6) {$x_2$};
	\node[my_obs] (x3) at (3,8) {$x_3$};
	\node[my_obs] (x4) at (6,8) {$x_4$};
	\node[my_obs] (x5) at (7,8) {$x_5$};

	\node[my_latent] (eta1) at (2,9) {$\eta_1$};
	\node[my_latent] (eta2) at (2,8) {$\eta_2$};
	\node[my_latent] (eta3) at (3,6.5) {$\eta_3$};
	\node[my_latent] (eta4) at (4.5, 8) {$\eta_4$};

	\draw[my_arrow] (eta1) -- (x3); 
	\draw[my_arrow] (eta2) -- (x3); 
	\draw[my_arrow] (x1) -- (x3); 
	\draw[my_arrow] (x1) -- (eta3); 
	\draw[my_arrow] (x2) -- (eta3); 
	\draw[my_arrow] (x3) -- (eta4); 
	\draw[my_arrow] (eta4) -- (x4); 
	\draw[my_arrow] (x4) -- (x5); 
	\draw[my_arrow] (x4) to [out=135, in=0] (4.5, 8.7) to [out=180, in=45]  (x3); 
	\draw[my_covariance] (eta2) to [out= 200, in=90] (1.2, 7) to [out=-90, in= 180]  (x2);
	\draw[my_covariance_mp] (y5) to [out= 0, in=225] (4.8, 6.2) to [out=45, in= -90]  (y6);
	
	\draw[my_arrow_mp] (eta1) -- (y1); 
	\draw[my_arrow_mp] (eta1) -- (y2); 
	\draw[my_arrow_mp] (eta1) -- (y3); 
	\draw[my_arrow_mp] (eta2) -- (y3); 
	\draw[my_arrow_mp] (eta3) -- (y4);
	\draw[my_arrow_mp] (eta3) -- (y5); 
	\draw[my_arrow_mp] (eta4) -- (y4);
	\draw[my_arrow_mp] (eta4) -- (y6);

	
	\end{scope}
	\end{tikzpicture}
	\caption{Example model}
	\label{fig:example}
\end{figure}
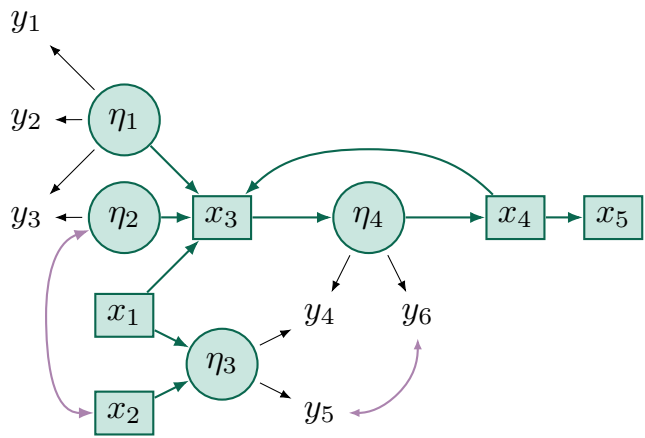

For example, let a linear equation in the structural part of SEM model take the form:
\[ \eta_3 = \beta_1 x_1 + \beta_2 x_2 + \varepsilon, \]
where $\eta_3$ is a variable dependent on regressors $x_1$ and $x_2$ (Fig. \ref{fig:example} ), $\beta_1$ and $\beta_2$ are parameters, $\varepsilon$ is an error term. In \textbf{semopy} syntax it can be rewritten as
\begin{lstlisting}[style=mycodestyle]
eta3 ~ x1 + x2
\end{lstlisting}


Likewise, to specify a measurement part, which relates manifest variables to latent variables, we use a special operator \lstinline{=~}  which can be read as \textit{is measured by}. The left side of the operator contains one latent variable, and the right contains its manifest variables separated by plus signs. For example, to define a latent variable $\eta_1$ by three indicators ($y_1, y_2$ and $y_3$) (Fig. \ref{fig:example} ), the following expression should be used :

\begin{lstlisting}[style=mycodestyle]
eta1 =~ y1 + y2 + y3
\end{lstlisting}

The third operator, \lstinline{~~}, is used to specify a covariance (common variance) between a pair of variables from one part:

\begin{lstlisting}[style=mycodestyle]
x1 ~~ x2
eta ~~ x3
\end{lstlisting}

Another option in the \textbf{semopy} syntax is fixing parameter values; the values should be placed as prefixes before variables:



\begin{lstlisting}[style=mycodestyle]
eta ~ 1*x1 + x2 
eta =~ 2*y1 + y2 + y3 
x1 ~~ 5*x2
\end{lstlisting}

We designed an example (Fig. \ref{fig:example}) to demonstrate the diversity of relationships between variables that can be estimated in \textbf{semopy}. The example contains two exogenous latent variable $\eta_1, \eta_2$, endogenous latent variables $\eta_3, \eta_4$ ($\eta_3$ being also an output variable), exogenous observed variables $x_1, x_2$, endogenous observed variables $x_3, x_4, x_5$ (the latter being an output variable) and a set of manifest variables $y_1, y_2, y_3, y_4, y_5, y_6$ (take a notice that $y_3$  and $y_4$ are shared between $\eta_1, \eta_2$ and $\eta_3, \eta_4$ respectively). There is also a cycle present in the model ($x_3 \rightarrow \eta_4 \rightarrow x_4 \rightarrow x_3$). Moreover, we set additional parameters for covariances between [$\eta_2$ and $x_2$] and between [$y_6$ and $y_5$].

The description of the model in the \textbf{semopy} syntax is as follows:


\begin{lstlisting}[style=mycodestyle]
# Structural part
eta3 ~ x1 + x2
eta4 ~ x3
x3 ~ eta1 + eta2 + x1 + x4
x4 ~ eta4
x5 ~ x4
# Measurement part
eta1 =~ y1 + y2 + y3
eta2 =~ y3
eta3 =~ y4 + y5
eta4 =~ y4 + y6
# Additional covariances
eta2 ~~ x2
y5 ~~ y6
\end{lstlisting}

%
%

In the \textbf{semopy}, we introduced an additional operator, that specifies a statistical type of variables. The "is" word is reserved  as an operator symbol for type specification:
\begin{lstlisting}[style=mycodestyle]
y1, y2 is ordinal
\end{lstlisting}
The first line sets $y_1, y_2$ to ordinal type.

\subsection{Quickstart}
The \textbf{semopy} package is available at PyPi software repository and can be installed by:
\begin{lstlisting}[style=mycodestyle]
pip install semopy
\end{lstlisting}

The pipeline for working with SEM models in \textbf{semopy} consists of three steps: (i) specifying a model, (ii) loading a dataset to the model, (iii) estimating parameters of the model. Two main objects required for scpecifying and estimating an SEM model are \colorbox{codecolors_inline}{\lstinline{Model}} and \colorbox{codecolors_inline}{\lstinline{Optimizer}}. 

\colorbox{codecolors_inline}{\lstinline{Model}} is responsible for setting up a model from the proposed SEM syntax:
\begin{lstlisting}[style=mycodestyle]
# The first step
from semopy import Model
mod = """ x1 ~ x2 + x3
          x3 ~ x2 + eta1
          eta1 =~ y1 + y2 + y3
          eta1 ~ x1
      """
model = Model(mod)
\end{lstlisting}
Then a dataset should be provided; at this step the initial values of parameters are calculated:
\begin{lstlisting}[style=mycodestyle]
# The second step
from pandas import read_csv
data = read_csv("my_data_file.csv", index_col=0)
model.load_dataset(data)
\end{lstlisting}

To estimate parameters of the model an \colorbox{codecolors_inline}{\lstinline{Optimizer}} object should be initialised and estimation executed:
\begin{lstlisting}[style=mycodestyle]
# The third step
from semopy import Optimizer
opt = Optimizer(model)
objective_function_value = opt.optimize()
\end{lstlisting}

The default objective function for estimating parameters is the likelihood function and the optimisation method is SLSQP (Sequential Least-Squares Quadratic Programming). However, the \textbf{semopy} supports a wide range of other objective functions and optimisation schemes being specified as parameters in the \colorbox{codecolors_inline}{\lstinline{optimize}} method (see Section \ref{sectopt}). 

%
%

When optimization process is finished, user can check the parameters' estimates and p-values by using \colorbox{codecolors_inline}{\lstinline{inspect}} method:

\begin{lstlisting}[style=mycodestyle]
from semopy import inspect
inspect(opt)
\end{lstlisting}

The important feature of the \textbf{semopy} is that one can run multiple optimization sessions in a row with preservation of previous parameters' estimates. 
For instance, to use unweighted least squares estimates as a starting point for a maximum likelihood estimation, one can run: 
\begin{lstlisting}[style=mycodestyle]
model = Model(mod)
model.load_dataset(data)
opt = Optimizer(model)
opt.optimize(objective='ULS')
opt.optimize(objective='MLW')
\end{lstlisting}

\section{Materials and methods}
In this section, we explain in detail the underlying calculations in the \textbf{semopy}. We denote latent variables with $\eta$, observed variables participating in relationships with latent variables with $x$ and manifest variables with $y$; let numbers of these variables be $n_\eta$, $n_x$ and $n_y$, respectively. We also assumed that all variables are normally distributed with zero means. Each group of variables, $\eta$ and $x$, can be split into two categories: (1) exogenous and (2) endogenous; let denote them $\eta^{(1)}$ and $\eta^{(2)}$, $x^{(1)}$ and $x^{(2)}$, respectively. Let the number of variables in the obtained four groups are $n_{\eta^{(1)}}$ and $n_{\eta^{(2)}}$, $n_{x^{(1)}}$ and $n_{x^{(2)}}$, so that $n_\eta = n_{\eta^{(1)}} + n_{\eta^{(2)}}$,  $n_x = n_{x^{(1)}} + n_{x^{(2)}}$.

\subsection{Model}
As usual for SEM models, we assume it consisted of two parts: structural and measurement ones. We consider the following generalisation of Eq. \ref{lisrel}:

\begin{equation} \label{eq:sem_init}
\begin{cases}

\begin{bmatrix} \eta \\ x  \end{bmatrix} 
&= \mathrm{B} \begin{bmatrix} \eta \\ x  \end{bmatrix} + \varepsilon \\
\hfil y &= \tilde{\Lambda} \eta + \delta
\end{cases},
\end{equation}
where $\tilde{\Lambda}$ is a factor loading matrix with linear parameters.
We introduce the following change of variables:
\[ \omega=\begin{bmatrix} \eta \\ x  \end{bmatrix} ; z = \begin{bmatrix} y \\ x  \end{bmatrix},\]
and the Eq. \ref{eq:sem_init} can be rewritten as:

\begin{equation} \label{eq:sem_our}
\begin{cases}
\omega &= \mathrm{B} \omega + \varepsilon \\
\hfil z &= \Lambda \omega + \delta
\end{cases},
\end{equation}
where $\mathrm{B}$ and $\Lambda$ are matrices of parameters, $\varepsilon$ and $\delta$ are vectors of error terms, which are assumed to be independent and normally distributed with zero means and covariances $\Psi$ and $\Theta$, respectively.

Then we can infer covariance matrix of $z$ as a function of  model parameters:

\begin{equation}\label{eq:cov}
cov(z) = \Sigma(\theta)
\end{equation}

%
%

\subsubsection*{Parameters in $\mathrm{B}$ matrix}

The size of $\mathrm{B}$ matrix corresponds to the size of $\omega$ vector and equals to $(n_\eta + n_x \times n_\eta + n_x)$. The block representation of this matrix is not necessary, and we set the initial values of all parameters in $\mathrm{B}$ matrix as zeroes.

\subsubsection*{Parameters in $\Lambda$ matrix}

The size of $\Lambda$ matrix matches to the sizes of $z$ and $\omega$ vectors and equals to $(n_y + n_x \times n_\eta + n_x)$; the block representation of $\Lambda$ is:

\renewcommand\arraystretch{1.5}
\begin{equation*}
\Lambda = \left[\begin{matrix}\tilde{\Lambda}(n_y\times n_{\eta })&0(n_y\times n_x)\\0(n_x\times n_{\eta })& \mathrm{I}(n_x\times n_x)\end{matrix}\right], 
\end{equation*} 
$\tilde{\Lambda}$ is an adjacency matrix for variables in a measurement part (Eq. \ref{eq:sem_init}).

To define the scale of latent variables, we automatically fix one parameter in each column of $\tilde{\Lambda}$  to 1. To be specific, in a column $i$ this parameter is the factor loading for $\eta_i$ latent variable and it's a manifest variable which is the first in the alphabet order of all manifest variable for $\eta_i$; we called this variable as the first indicator. Initial values of the remaining factor loading for the $\eta_i$ are linear regression coefficients between manifest variables and the first indicator being as a single regressor. 

\subsubsection*{Parameters in $\Psi$ matrix}

The matrix $\Psi$ is a square covariance matrix for the error term $\varepsilon$ (Eq. \ref{eq:sem_our}) and it's size is $(n_\eta + n_x \times n_\eta + n_x)$ or, in other terms, $(n_{\eta^{(1)}} + n_{\eta^{(2)}} + n_{x^{(1)}} + n_{x^{(2)}} \times n_{\eta^{(1)}} + n_{\eta^{(2)}} + n_{x^{(1)}} + n_{x^{(2)}})$. This matrix is symmetric and can be presented in the block form:

\renewcommand\arraystretch{1.5}
\begin{equation*}
\Psi = \left[\begin{matrix}

\Psi _{\eta^{(1)}} & \Psi _{\eta^{(1)}, \eta^{(2)}} 
& \Psi _{\eta^{(1)}, x^{(1)}} & \Psi _{\eta^{(1)}, x^{(2)}}
\\
\Psi _{\eta^{(1)}, \eta^{(2)}}^\intercal & \Psi _{\eta^{(2)}} 
& \Psi _{\eta^{(2)}, x^{(1)}} & \Psi _{\eta^{(2)}, x^{(2)}}
\\
\Psi _{\eta^{(1)}, x^{(1)}}^\intercal & \Psi _{\eta^{(2)}, x^{(1)}}^\intercal 
& \Psi _{x^{(1)}} & \Psi _{x^{(1)}, x^{(2)}}
\\
\Psi _{\eta^{(1)}, x^{(2)}}^\intercal & \Psi _{\eta^{(2)}, x^{(2)}}^\intercal 
& \Psi _{x^{(1)}, x^{(2)}}^\intercal & \Psi _{x^{(2)}}

\end{matrix}\right],
\end{equation*}
where blocks reflect covariance matrices of variables mentioned in indexes. By default, we assume that the block $\Psi _{x^{(1)}}$ is fixed and equals to sample covariance matrix for $x^{(1)}$ variables (exogenous observed variables). For $\eta^{(1)}$ (exogenous latent variables), we assumed the symmetric $\Psi _{\eta^{(1)}}$ covariance matrix fully parametrised. Preventing covariances between $x^{(1)}$ and $\eta^{(1)}$ by default, we set $\Psi _{\eta^{(1)},x^{(1)}}$ as zero  matrix. We also consider the covariance matrices between endogenous and exogenous variables as zero matrices. In the remaining matrices for covariances between endogenous variables (latent and observed) -- $\Psi _{\eta^{(2)}}$, $ \Psi _{x^{(2)}}$, $\Psi _{\eta^{(2)}, x^{(2)}}$ -- we set parameters in positions of variances and in positions of covariances between variables, which do not play a role of regressors in any equation of the structural part.

%

\subsubsection*{Parameters in Theta matrix}

The $\Theta$ matrix is symmetric square 4-block matrix of $(n_y + n_x \times n_y + n_x)$ size having only one non-zero block ($\tilde{\Theta}$):

\renewcommand\arraystretch{1.5}
\begin{equation*}
\Theta = \left[\begin{matrix}\tilde{\Theta}(n_y\times n_y) & 0(n_y\times n_x) 
\\
0(n_x\times n_y)&0(n_x\times n_x)\end{matrix}\right].
\end{equation*} 

Dy default we initialise $\tilde{\Theta}$ as a diagonal matrix, however, the \textbf{semopy} syntax allows to parametrise it's off-diagonal elements. We set the starting values for a diagonal element $\tilde{\Theta}_{i,i}$ as a half of a sample variance for the manifest variable $y_i$.



\subsubsection*{Example}
\begin{figure}[H] 
	\centering
	\begin{tabular}{cc}
	$\mathrm{B}$ matrix & $\Psi$ matrix \\
	\begin{tikzpicture}[font=\sffamily\huge]
	\begin{scope}[scale=0.4, transform shape]

	\draw[color=black, fill=gray_box, step=1cm] (1, -1) rectangle(10, 8);

	\draw[color=g_box, fill=g_box, step=1cm](2, 4) rectangle(1, 3);
	\draw[color=g_box, fill=g_box, step=1cm](2, 4) rectangle(3, 3);
	\draw[color=g_box, fill=g_box, step=1cm](5, 3) rectangle(4, 2);
	\draw[color=g_box, fill=g_box, step=1cm](6, 5) rectangle(5, 4);
	\draw[color=g_box, fill=g_box, step=1cm](7, 2) rectangle(6, 1);
	\draw[color=g_box, fill=g_box, step=1cm](7, 4) rectangle(6, 3);
	
	\draw[color=g_box, fill=g_box, step=1cm](9, 6) rectangle(8, 5);
	\draw[color=g_box, fill=g_box, step=1cm](9, 4) rectangle(8, 3);
	
	\draw[color=g_box, fill=g_box, step=1cm](10, 6) rectangle(9, 5);

	\draw[step=1cm, color=black] (0, -1) grid (10,9);
	\node at (1.5, 8.5) {$\eta_1$};
	\node at (2.5, 8.5) {$\eta_2$};
	\node at (3.5, 8.5) {$\eta_3$};
	\node at (4.5, 8.5) {$\eta_4$};
	\node at (5.5, 8.5) {$x_3$};
	\node at (6.5, 8.5) {$x_4$};
	\node at (7.5, 8.5) {$x_5$};
	\node at (8.5, 8.5) {$x_1$};
	\node at (9.5, 8.5) {$x_2$};
	
	\node at (0.5, 7.5) {$\eta_1$};
	\node at (0.5, 6.5) {$\eta_2$};
	\node at (0.5, 5.5) {$\eta_3$};
	\node at (0.5, 4.5) {$\eta_4$};
	\node at (0.5, 3.5) {$x_3$};
	\node at (0.5, 2.5) {$x_4$};
	\node at (0.5, 1.5) {$x_5$};
	\node at (0.5, 0.5) {$x_1$};
	\node at (0.5, -0.5) {$x_2$};
	\end{scope}
	\end{tikzpicture}
	&
	
	\begin{tikzpicture}[font=\sffamily\huge]
	\begin{scope}[scale=0.4, transform shape]
	\draw[color=black, fill=gray_box, step=1cm] (1, -1) rectangle(10, 8);

	\draw[color=g_box, fill=g_box, step=1cm](2, 8) rectangle(1, 7);
	\draw[color=g_box, fill=g_box, step=1cm](3, 0) rectangle(2, -1);
	\draw[color=g_box, fill=g_box, step=1cm](8, 1) rectangle(7, 2);
	\draw[color=g_box, fill=g_box, step=1cm](4, 2) rectangle(3, 1);
	\draw[color=g_box, fill=g_box, step=1cm](7, 5) rectangle(8, 6);
	\draw[color=g_box, fill=g_box, step=1cm](10, 7) rectangle(9, 6);
	\draw[color=g_box, fill=g_box, step=1cm](3, 8) rectangle(2, 7);
	\draw[color=g_box, fill=g_box, step=1cm](2, 7) rectangle(1, 6);
	\draw[color=g_box, fill=g_box, step=1cm](3, 7) rectangle(2, 6);
	\draw[color=g_box, fill=g_box, step=1cm](4, 6) rectangle(3, 5);
	\draw[color=g_box, fill=g_box, step=1cm](5, 5) rectangle(4, 4);
	\draw[color=g_box, fill=g_box, step=1cm](6, 4) rectangle(5, 3);
	\draw[color=g_box, fill=g_box, step=1cm](7, 3) rectangle(6, 2);
	
	\draw[color=y_box, fill=y_box, step=1cm](10,1) rectangle(8, -1);
	\draw[step=1cm, color=black] (0, -1) grid (10,9);
	\node at (1.5, 8.5) {$\eta_1$};
\node at (2.5, 8.5) {$\eta_2$};
\node at (3.5, 8.5) {$\eta_3$};
\node at (4.5, 8.5) {$\eta_4$};
\node at (5.5, 8.5) {$x_3$};
\node at (6.5, 8.5) {$x_4$};
\node at (7.5, 8.5) {$x_5$};
\node at (8.5, 8.5) {$x_1$};
\node at (9.5, 8.5) {$x_2$};

\node at (0.5, 7.5) {$\eta_1$};
\node at (0.5, 6.5) {$\eta_2$};
\node at (0.5, 5.5) {$\eta_3$};
\node at (0.5, 4.5) {$\eta_4$};
\node at (0.5, 3.5) {$x_3$};
\node at (0.5, 2.5) {$x_4$};
\node at (0.5, 1.5) {$x_5$};
\node at (0.5, 0.5) {$x_1$};
\node at (0.5, -0.5) {$x_2$};
	\end{scope}
	\end{tikzpicture} 
	\\

	$\Lambda$ matrix & $\Theta$ matrix \\
	
	\begin{tikzpicture}[font=\sffamily\huge]
	\begin{scope}[scale=0.4, transform shape]
	\draw[color=black, fill=gray_box, step=1cm] (1, -3) rectangle(10, 8);

	\draw[color=g_box, fill=g_box, step=1cm](2, 7) rectangle(1, 6);
	
	
		\node at (1.5, 7.5) {1};
	\draw[color=g_box, fill=g_box, step=1cm](2, 7) rectangle(1, 6);
	\draw[color=g_box, fill=g_box, step=1cm](2, 6) rectangle(1, 5);
	\node at (2.5, 5.5) {1};
	\node at (3.5, 4.5) {1};
	\draw[color=g_box, fill=g_box, step=1cm](3, 3) rectangle(4, 4);
	\node at (4.5, 4.5) {1};
	\draw[color=g_box, fill=g_box, step=1cm](4, 2) rectangle(5, 3);
	\node at (5.5,  1.5) {1};
	\node at (6.5,  0.5) {1};
	\node at (7.5,  -0.5) {1};
	\node at (8.5,  -1.5) {1};
	\node at (9.5,  -2.5) {1};
	\draw[step=1cm, color=black] (0, -3) grid (10, 9);
	\node at (1.5, 8.5) {$\eta_1$};
	\node at (2.5, 8.5) {$\eta_2$};
	\node at (3.5, 8.5) {$\eta_3$};
	\node at (4.5, 8.5) {$\eta_4$};
	\node at (5.5, 8.5) {$x_3$};
	\node at (6.5, 8.5) {$x_4$};
	 \node at (7.5, 8.5) {$x_5$};
	 \node at (8.5, 8.5) {$x_1$};
	  \node at (9.5, 8.5) {$x_2$};
	\node at (0.5, 7.5) {$y_1$};
	\node at (0.5, 6.5) {$y_2$};
	\node at (0.5, 5.5) {$y_3$};
	\node at (0.5, 4.5) {$y_4$};
	 \node at (0.5, 3.5) {$y_5$};
	 \node at (0.5, 2.5) {$y_6$};
	 \node at (0.5, 1.5) {$x_3$};
	 \node at (0.5, 0.5) {$x_4$};
	  \node at (0.5, -0.5) {$x_5$};
	  \node at (0.5, -1.5) {$x_1$};
	  \node at (0.5, -2.5) {$x_2$};

	\end{scope}
	\end{tikzpicture} 
	&
	
		\begin{tikzpicture}[font=\sffamily\huge]
	\begin{scope}[scale=0.4, transform shape]
	\draw[color=black, fill=gray_box, step=1cm] (1, -3) rectangle(12, 8);
\draw[color=g_box, fill=g_box, step=1cm](2, 8) rectangle(1, 7);
\draw[color=g_box, fill=g_box, step=1cm](3, 7) rectangle(2, 6);
\draw[color=g_box, fill=g_box, step=1cm](4, 6) rectangle(3, 5);
\draw[color=g_box, fill=g_box, step=1cm](5, 5) rectangle(4, 4);
\draw[color=g_box, fill=g_box, step=1cm](6, 4) rectangle(5, 3);
\draw[color=g_box, fill=g_box, step=1cm](7, 3) rectangle(6, 2);

\draw[color=g_box, fill=g_box, step=1cm](7, 4) rectangle(6, 3);
\draw[color=g_box, fill=g_box, step=1cm](6, 3) rectangle(5, 2);
\draw[step=1cm, color=black] (0, -3) grid (12,9);
\node at (1.5, 8.5) {$y_1$};
\node at (2.5, 8.5) {$y_2$};
\node at (3.5, 8.5) {$y_3$};
\node at (4.5, 8.5) {$y_4$};
\node at (5.5, 8.5) {$y_5$};
\node at (6.5, 8.5) {$y_6$};
\node at (7.5, 8.5) {$x_3$};
\node at (8.5, 8.5) {$x_4$};
\node at (9.5, 8.5) {$x_5$};
\node at (10.5, 8.5) {$x_1$};
\node at (11.5, 8.5) {$x_2$};
\node at (0.5, 7.5) {$y_1$};
\node at (0.5, 6.5) {$y_2$};
\node at (0.5, 5.5) {$y_3$};
\node at (0.5, 4.5) {$y_4$};
\node at (0.5, 3.5) {$y_5$};
\node at (0.5, 2.5) {$y_6$};
\node at (0.5, 1.5) {$x_3$};
\node at (0.5, 0.5) {$x_4$};
\node at (0.5, -0.5) {$x_5$};
\node at (0.5, -1.5) {$x_1$};
\node at (0.5, -2.5) {$x_2$};
	\end{scope} 
	\end{tikzpicture}
	\end{tabular}
\caption{Matrices for model on Fig. \ref{fig:example}. Gray cells stand for zeros, green cells stand for parameters, purple cells stand for values from sample covariance matrix.}
\label{fig:matrices}
\end{figure}
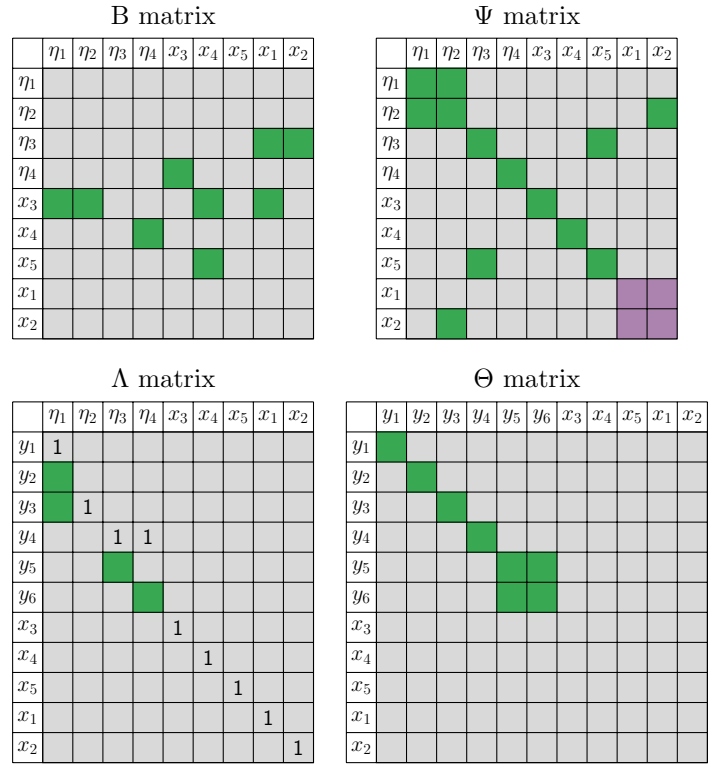

For the model on Fig. \ref{fig:example}, positions of parameters in matrices are presented on Fig. \ref{fig:matrices}.

\subsection{Loading the data}

The \textbf{semopy} supports SEM models that contain not only continuous variables (assumed as normally distributed) but also discrete ones (assumed as ordinal). By default, all variables are assumed to be normally distributed. However, it is possible to manually specify the type of variables or to allow the automatic recognition of types. 

\begin{lstlisting}[style=mycodestyle]
# Providing a list of ordinal variables
variables = {'y1', 'y2'}
model.load_dataset(data, ordcor=variables)

# Load data with automatically recognition of types
model.load_dataset(data, ordcor=True)

# Check set of ordinal variables
print(model.vars['Categorical'])
\end{lstlisting}


Presence of ordinal variables makes the sample covariance matrix $S$ "heterogenous", i.e. containing polychoric (in-between ordinal) and polyserial (between ordinal and continuous variables) correlations \cite{polycorr} .


\subsection{Objective functions}
The \textbf{semopy} allows to chose one of three objective (loss) functions for parameters' estimation. Two of them are based on a least-squares approach, and the remaining one (the default) represents the maximum likelihood approach. It should be noticed that all objective functions reflect (in different ways) the distance between the sample covariance matrix $\mathrm{S}$ and the model covariance matrix $\Sigma(\theta)$ for observed variables $z$. In this section, we present both objective functions and their two first derivatives utilised in optimisation methods.

The objective function can be set as an argument in \colorbox{codecolors_inline}{\lstinline{optimize}} method of \colorbox{codecolors_inline}{\lstinline{Optimizer}} class:
\begin{lstlisting}[style=mycodestyle]
opt.optimize(objective='ULS') # Unweighted least squares
opt.optimize(objective='GLS') # General least squares
opt.optimize(objective='MLW') # Wishart likelihood function
\end{lstlisting}

\subsubsection*{Unweighted Least Squares (ULS)}
The ULS objective function can be written as follows:

\begin{equation}
F_{ULS}(\theta ) = \mathrm{tr} \left[(\Sigma(\theta) - S)(\Sigma(\theta)-S)^\intercal \right],
\end{equation}
where $\theta$ is a set of all parameters in an SEM model. To accelerate some optimisation methods, we inferred formulas for components in the gradient and the Hessian for $F_{ULS}(\theta )$.

\subsubsection*{General Least Squares (GLS)}
In contrast to ULS loss function, GLS approach considers the following loss function:

\begin{equation}
F(\theta )_{GLS} = \mathrm{tr}\left[(E-\Sigma(\theta) S^{-1})^2\right]
\end{equation}

After our inference, formulas for components in the gradient and the Hessian are available as well.

\subsubsection*{Wishart Maximum Likelihood (MLW)} \label{mlw}
The MLW objective function is based on the assumption that the observed variables $z$ follow the multivariate normal distribution, therefore, the sample covariance matrix of $z$ follows the Wishart distribution:

\begin{equation*}
\centering
W(S| n^{-1}\Sigma, n) \propto \frac{e^{- \mathrm{tr} (nS\Sigma^{-1})/2}|nS|^{(n - p - 1)/2} }{|\Sigma|^{n/2}} \propto e^{-\mathrm{tr}(nS \Sigma^{-1})/2} \left| \Sigma\right|^{-n/2},
\end{equation*}
where $n$ is number of degrees of freedom (sample size), $p$ is number of parameters ($=|\theta|$). In this case, the log likelihood ratio -- a natural logarithm of ratio of likelihood for any given model to likelihood with a perfectly fitting model ($\Sigma = S$) -- is following:

\begin{equation}\label{eq:mlw}
F_{MLW}(\theta )\;=\;\mathrm{tr}\left[S\Sigma(\theta) ^{-1}\right]+\ln \left|\Sigma(\theta) \right|
\end{equation}

Also, we inferred the analytical gradient and Hessian of  $F_{MLW}$.

\subsection{Optimisation methods}
\label{sectopt}
To minimise an objective function, \textbf{semopy} has a variety of nonlinear solvers. Optimisation method can be selected through \colorbox{codecolors_inline}{\lstinline{method}} argument to \colorbox{codecolors_inline}{\lstinline{optimize}} function in \colorbox{codecolors_inline}{\lstinline{Optimizer}} class, for example, 
\begin{lstlisting}[style=mycodestyle]
opt.optimize(method="SLSQP")
\end{lstlisting}

The full list of available optimisation methods is the following:

\begin{itemize}
	\item \colorbox{codecolors_inline}{\lstinline{SLSQP}} -- SLSQP (Sequential Least-Squares Quadratic Programming) method \cite{kraftd.1988, kraftd.1994} from \textbf{scipy};
	\item \colorbox{codecolors_inline}{\lstinline{L-BFGS-B}} -- L-BFGS-B (Broyden — Fletcher — Goldfarb — Shanno, limited memory) method \cite{numopt.wright, zhu1997} from \textbf{scipy};
	\item  \colorbox{codecolors_inline}{\lstinline{Portmin}} -- FORTRAN \textbf{PORT} optimization library \cite{gay.1983} wrapped with Python portmin wrapper. It incorporates \textit{SMSNO, SUMSL, HUMSL} routines for cases when no analytical gradient is available, when analytical gradient is available and when both analytical gradient and hessian are available respectively;
	\item  \colorbox{codecolors_inline}{\lstinline{Adam}} -- Stochastic optimisation method Adam\cite{adam, ruder2016}, our implementation;
	\item  \colorbox{codecolors_inline}{\lstinline{Nesterov}} --Stochastic Nesterov Accelerated Gradient method \cite{ruder2016}, our implementation;
	\item  \colorbox{codecolors_inline}{\lstinline{SGD}} -- Stochastic Gradient Descent method \cite{ruder2016}, our implementation.
\end{itemize}


\subsection{Statistics}

The \textbf{semopy} provides methods to calculate important statistics: p-values for parameter estimates and various measurements of fit.


\subsubsection*{P-values for parameter estimates}

The \textbf{semopy} utilises the Z-test to calculate p-values for parameter estimates under the assumption that parameters are normality distributed; $H_0$: value of a parameter is equal to zero. The approach considers z-score:
\[ Z(\hat{\theta}) = \frac{\hat{\theta}}{SE(\hat{\theta}) }, \]
where $\hat{\theta}$ is a vector of parameter estimates, $SE(\hat{\theta})$ is the standard error of estimates, which is proportional to variance: $SE(\hat{\theta}) = var(\hat{\theta})/\sqrt{n}$.

Based on the Cramér–Rao bound, 
$$var(\hat{\theta}) \ge \mathrm{FIM}(\theta)^{-1}, $$
where $\mathrm{FIM}(\theta)$ is a Fisher information matrix (FIM). The $\mathrm{FIM}(\theta)$ matrix can be defined as an observed or expected FIM. The observed FIM is the Hessian of a likelihood function at $\hat{\theta}$, $H(\hat{\theta})$.

P-values for parameter estimates are based on the Z-test expectation, that  $Z(\theta)$ follows the multivariate normal distribution. By default, expected FIM is used, but a user may want to estimate observed FIM instead:
\begin{lstlisting}[style=mycodestyle]
from semopy.stats import calculate_p_values
pvals = calculate_p_values(opt, information='observed')
\end{lstlisting}

\subsubsection*{Fit indices}

The \textbf{semopy} supports numerous fit indices: $\chi^2$, RMSEA, CFI, TLI, NFI GFI, AGFI \cite{kline.2015, hufit}. Some fit indices compare fitted and baseline (null or independence) models. In \textbf{semopy}, baseline model is a model with all regression coefficients, loading factors, covariances, variances of latent variables set to zero. Several methods use the degrees of freedom ($df$) metric, $df = \frac{k (k + 1)}{2} - m$, where $k$ is a number of observed variables and $m$ is a number of parameters.
All methods that estimate them require an \colorbox{codecolors_inline}{\lstinline{Optimizer}} instance as an argument.
\begin{table}[H]
	\centering
	\caption{where $n$ is a number of data samples, $F(\hat{\theta})$ is a value that objective function attains at optimum, $\chi^2_m$ is a $\chi^2$ statistics for the target model, $\chi^2_b$ is a $\chi^2$ statistics for the baseline model, where $df_m$ is $df$ of target model and $df_b$ is $df$ of baseline model, $k$ is a number of parameters and $L$ is a value of a likelihood function.}
	\begin{tabular}{|c|c|l|l|}
		\hline
		Fit index & Formula & Method \\\hline
		$\chi^2$ & $ n F(\hat{\theta})$ & \colorbox{codecolors_inline}{\lstinline{calc\_chi2}} \\\hline
		RMSEA & $ \sqrt{\frac{\chi^2 / df - 1}{n - 1}}$ & \colorbox{codecolors_inline}{\lstinline{calc\_rmsea}} \\\hline
		GFI & $1 - \frac{\chi^2_m}{\chi^2_b}$ & \colorbox{codecolors_inline}{\lstinline{calc\_gfi}} \\\hline
		AGFI & $ 1 - \frac{k ( k + 1)} {2 df} (1 - GFI)$ & \colorbox{codecolors_inline}{\lstinline{calc\_agfi}} \\\hline
		NFI & $\frac{\chi^2_b - \chi^2_m}{\chi^2_b}$ & \colorbox{codecolors_inline}{\lstinline{calc\_nfi}} \\\hline
		TLI & $\frac{\frac{\chi^2_b}{df_b} - \frac{\chi^2_m}{df_m}}{\frac{\chi^2_b}{df_b} - 1}$ & \colorbox{codecolors_inline}{\lstinline{calc\_tli}} \\\hline
		CFI & $1 - \frac{\chi^2_m - df_m}{\chi^2_b - df_b} $ & \colorbox{codecolors_inline}{\lstinline{calc\_cfi}} \\\hline
		AIC & $2(k - L) $ & \colorbox{codecolors_inline}{\lstinline{calc\_aic}} \\\hline
		BIC & $\mathbf(ln)(n) k - 2  L$ & \colorbox{codecolors_inline}{\lstinline{calc\_bic}} \\\hline
\end{tabular}

\end{table}
A fit index can be calculated by invoking a particular method with an \colorbox{codecolors_inline}{\lstinline{Optimizer}} instance passed as an argument. For example, to calculate AIC or BIC one should invoke \colorbox{codecolors_inline}{\lstinline{calculate\_aic}} or \colorbox{codecolors_inline}{\lstinline{calculate\_bic}} method respectively. They  automatically estimate $L$ as a Wishart likelihood by default. However, one might supply extra parameter \textit{lh} as a value of $L$: 
\begin{lstlisting}[style=mycodestyle]
from semopy import calc_aic, calc_likelihood
# Calculate multivariate normal likelihood of model fitted by opt
l = calc_likelihood(opt, dist='normal') 
aic = calc_aic(opt, lh=l)
\end{lstlisting}

Alternatively, one might gather all statistics explained above by a single call to \colorbox{codecolors_inline}{\lstinline{gather\_statistics}} method:
\begin{lstlisting}[style=mycodestyle]
from semopy import gather_statistics
stats = gather_statistics(opt)
\end{lstlisting}

\subsection{Testing framework}
In order to test the \textbf{semopy} package and compare it's performance with the \textbf{lavaan}, we implemented a versatile system to generate benchmark SEM models. Based on input parameters, the system randomly generates a skeleton of structural and measurement parts of an SEM model, values of all parameters and the dataset appropriate for the model and parameters. The set of generator's parameters includes:
\begin{itemize}
	\item \colorbox{codecolors_inline}{\lstinline{n_obs}} , total number of variables in the structural part;
	\item \colorbox{codecolors_inline}{\lstinline{n_lat}}, number of latent variables in the structural part;
	\item \colorbox{codecolors_inline}{\lstinline{n_cycle}}, minimal number of cycles in the structural part;
	\item \colorbox{codecolors_inline}{\lstinline{n_manif/(l_manif, u_manif)}}, range of possible numbers of manifest variables for a latent variable;
	\item \colorbox{codecolors_inline}{\lstinline{p_manif}}, fraction of manifest variables to merge together.
	\item \colorbox{codecolors_inline}{\lstinline{scale}} all parameters sampled from uniform distribution on domain $[-1.0, -0.1] \cup [0.1, 1.0]$ are multiplied by this value.
	\item \colorbox{codecolors_inline}{\lstinline{n_samples}} number of samples in a dataset.
\end{itemize}

In order to generate a model, one should run the following:

\begin{lstlisting}[style=mycodestyle]
from semopy.model_generator import generate_model

n_manif = (l_manif, u_manif)
model, params, data = generate_model(n_obs, n_lat, n_manif,
                                     p_manif, n_cycles, scale,
                                     n_samples)
\end{lstlisting}

The algorithm generating benchmarks consists of four steps. At the first step, we construct a structural part of an SEM model. For this purpose, we randomly generate a directed acyclic graph of (\colorbox{codecolors_inline}{\lstinline{n_obs}} + \colorbox{codecolors_inline}{\lstinline{n_lat}}) nodes and then we add \colorbox{codecolors_inline}{\lstinline{n_cycles}} extra directed edges between nodes to satisfy the minimal number of circles in the structural part. At last, \colorbox{codecolors_inline}{\lstinline{n_lat}} nodes in the structural parts are picked as latent. At the second step, we construct the measurement part providing each latent variable with it's own set of manifest variables, so that each latent variable has the random amount of manifest variables from the range \colorbox{codecolors_inline}{\lstinline{(l_manif, u_manif)}}. To allow a fraction of manifest variables measuring several latent variables, we consequently merge pairs of manifest variables from different latent variables until the resultant number of manifest variables reaches \colorbox{codecolors_inline}{\lstinline{(1-f_manif)}} of the initial number of manifest variables.
At the third step, values of parameters in $\mathrm{B}$ and $\Lambda$ are uniformly sampled from the continuous uniform distribution on the symmetric support $[-1, -0.1] \cup [0.1, 1]$ scaled by \colorbox{codecolors_inline}{\lstinline{scale}} parameter. At the fourth step, we consistently generate values for variables starting from exogenous ones, moving along paths in the structural part, finishing with manifest variables. For each variable we add a random noise error from $\mathcal{N}(0, 0.1)$. After a primary dataset is generated, we removed the latent variables form it.

%
%
%
%
%

\section{Results}

To compare \textbf{semopy} and \textbf{lavaan} packages to each other we tested them on 15 benchmark sets of SEM models generated by the model generator explained above. Sets consist of 1000 random models and different parameters (see Table \ref{tableModels}).

\begin{table}[H]
\centering
\caption{Parameters that were passed to generator for each set. Sets are separated into groups with a horizontal line paying respect to our representation of tests' results on graphs.  \colorbox{codecolors_inline}{\lstinline{p_manif}} is set to 0.1 everywhere.}
\label{tableModels}
\begin{adjustbox}{max width=\textwidth} 
\begin{tabular}{|c|c|c|c|c|c|c|}
\hline
N & n\_obs & n\_lat & n\_manif & n\_cycles & scale & n\_samples\\ 
\hline
1 & 5 & 2 & 2 & 0 & \textbf{0.5} & 500\\ 

2 & 5 & 2 & 2 & 0 & \textbf{0.75} & 500\\ 

3 & 5 & 2 & 2 & 0 & \textbf{1} & 500\\ 

4 & 5 & 2 & 2 & 0 & \textbf{1.5} & 500\\ 

5 & 5 & 2 & 2 & 0 & \textbf{2} & 500\\ 
\hline
6 & 10 & \textbf{0} & 2 & 0 & 1 & 500\\ 

7 & 10 & \textbf{1} & 2 & 0 & 1 & 500\\ 

8 & 10 & \textbf{2} & 2 & 0 & 1 & 500\\ 

9 & 10 & \textbf{4} & 2 & 0 & 1 & 500\\ 

10 & 10 & \textbf{8} & 2 & 0 & 1 & 500\\ 
\hline
11 & 10 & \textbf{0} & 2 & 1 & 1 & 500\\ 

12 & 10 & \textbf{1} & 2 & 1 & 1 & 500\\ 

13 & 10 & \textbf{2} & 2 & 1 & 1 & 500\\ 

14 & 10 & \textbf{4} & 2 & 1 & 1 & 500\\ 

15 & 10 & \textbf{8} & 2 & 1 & 1 & 500\\ 
\hline
\end{tabular}
\end{adjustbox}
\end{table}

Then, we ran \textbf{semopy} and \textbf{lavaan} on each model from those sets and compared packages on the following measures of accuracy:
\begin{itemize}
\item relative error between the estimated ($\hat{\theta}$) and exact parameter values ($\theta$). As $\theta$ is a vector, we consider a mean relative error  $\Delta(\theta, \hat{\theta}) = \frac{1}{n} \sum_{i=1}^n \frac{|\hat{\theta_i} - \theta_i |}{|\theta_i|}$;
\item the obtained value of the objective functions $F(\hat{\theta})$;
\item number of failed optimisation processes.
\item execution time
\end{itemize}


We consider an optimisation process failed (a package's estimates diverge from true values) if any of the criteria are met:
\begin{itemize}
	\item any parameter has a "Not-a-Number" (NaN) value;
	\item the objective function returns NaN at given point (function with estimated parameters attains undefined value and/or violates innerly-defined boundaries).
	\item $\Delta(\theta, \hat{\theta}) < 0.3$
\end{itemize} 

15 sets of models that we examined in our tests are provided in Table \ref{tableModels}. All sets of models and their respective data used in the following tests as well as the results are available at the repository.

Next, we provide the results of our tests. 
\begin{figure}[H]
	\centering
	\begin{adjustbox}{max width=\textwidth} 
	\begin{tabular}{ccc}
		MLW & ULS & GLS \\
		\begin{tikzpicture}
		\begin{scope}[scale=0.46, transform shape]
		\pgfkeys{{/pgfplots/legend entry/.code=\addlegendentry{#1}}}
		\begin{axis}[
		ylabel={$N$, number of non-convergent models},
		legend pos=north east,
		ymajorgrids=true,
		grid style=dashed,
		]
		\addplot[
		color=red,
		mark=square,
		legend entry=lavaan,
		only marks
		]
		coordinates {
			(1,282)(2,135)(3,45)(4,13)(5,5)
		};
		\addplot[
		color=blue,
		mark=o,
		legend entry=semopy,
		only marks
		]
		coordinates {
			(1,266)(2,123)(3,41)(4,13)(5,4)
		};
		\end{axis}
		\end{scope}
		\end{tikzpicture}
		&
		\begin{tikzpicture}
		\begin{scope}[scale=0.46, transform shape]
		\pgfkeys{{/pgfplots/legend entry/.code=\addlegendentry{#1}}}
		\begin{axis}[
		legend pos=north east,
		ymajorgrids=true,
		grid style=dashed,
		]
		\addplot[
		color=red,
		mark=square,
		only marks
		]
		coordinates {
			(1,367)(2,190)(3,81)(4,44)(5,77)
		};
		\addplot[
		color=blue,
		mark=o,
		only marks
		]
		coordinates {
			(1,366)(2,188)(3,76)(4,41)(5,65)
		};
		\end{axis}
		\end{scope}
		\end{tikzpicture}
		&
		\begin{tikzpicture}
		\begin{scope}[scale=0.46, transform shape]
		\pgfkeys{{/pgfplots/legend entry/.code=\addlegendentry{#1}}}
		\begin{axis}[
		legend pos=north east,
		ymajorgrids=true,
		grid style=dashed,
		]
		\addplot[
		color=red,
		mark=square,
		only marks
		]
		coordinates {
			(1,287)(2,158)(3,77)(4,56)(5,90)
		};
		\addplot[
		color=blue,
		mark=o,
		only marks
		]
		coordinates {
			(1,265)(2,139)(3,54)(4,42)(5,80)
		};
		\end{axis}
		\end{scope}
		\end{tikzpicture} \\
		(A) & (B) & (C) \\
		\begin{tikzpicture}
		\begin{scope}[scale=0.46, transform shape]
		\pgfkeys{{/pgfplots/legend entry/.code=\addlegendentry{#1}}}
		\begin{axis}[
		ylabel={$N$, number of non-convergent models},
		legend pos=north east,
		ymajorgrids=true,
		grid style=dashed,
		]
		\addplot[
		color=red,
		mark=square,
		only marks
		]
		coordinates {
			(6,33)(7,45)(8,48)(9,112)(10,197)
		};
		\addplot[
		color=blue,
		mark=o,
		only marks
		]
		coordinates {
			(6,33)(7,39)(8,43)(9,76)(10,133)
		};
		\end{axis}
		\end{scope}
		\end{tikzpicture}
		&
		\begin{tikzpicture}
		\begin{scope}[scale=0.46, transform shape]
		\pgfkeys{{/pgfplots/legend entry/.code=\addlegendentry{#1}}}
		\begin{axis}[
		legend pos=north east,
		ymajorgrids=true,
		grid style=dashed,
		]
		\addplot[
		color=red,
		mark=square,
		only marks
		]
		coordinates {
			(6,67)(7,77)(8,98)(9,181)(10,309)
		};
		\addplot[
		color=blue,
		mark=o,
		only marks
		]
		coordinates {
			(6,68)(7,78)(8,103)(9,178)(10,278)
		};
		\end{axis}
		\end{scope}
		\end{tikzpicture}
		&
		\begin{tikzpicture}
		\begin{scope}[scale=0.46, transform shape]
		\pgfkeys{{/pgfplots/legend entry/.code=\addlegendentry{#1}}}
		\begin{axis}[
		legend pos=north east,
		ymajorgrids=true,
		grid style=dashed,
		]
		\addplot[
		color=red,
		mark=square,
		only marks
		]
		coordinates {
			(6,48)(7,74)(8,74)(9,159)(10,285)
		};
		\addplot[
		color=blue,
		mark=o,
		only marks
		]
		coordinates {
			(6,44)(7,61)(8,46)(9,107)(10,217)
		};
		\end{axis}
		\end{scope}
		\end{tikzpicture}\\
		(D) & (E) & (F) \\
		\begin{tikzpicture}
		\begin{scope}[scale=0.46, transform shape]
		\pgfkeys{{/pgfplots/legend entry/.code=\addlegendentry{#1}}}
		\begin{axis}[
		xlabel={Set},
		ylabel={$N$, number of non-convergent models},
		legend pos=north east,
		ymajorgrids=true,
		grid style=dashed,
		]
		\addplot[
		color=red,
		mark=square,
		only marks
		]
		coordinates {
			(11,111)(12,144)(13,127)(14,212)(15,291)
		};
		\addplot[
		color=blue,
		mark=o,
		only marks
		]
		coordinates {
			(11,111)(12,143)(13,122)(14,197)(15,232)
		};
		\end{axis}
		\end{scope}
		\end{tikzpicture}
		&
		\begin{tikzpicture}
		\begin{scope}[scale=0.46, transform shape]
		\pgfkeys{{/pgfplots/legend entry/.code=\addlegendentry{#1}}}
		\begin{axis}[
		xlabel={Set},
		legend pos=north east,
		ymajorgrids=true,
		grid style=dashed,
		]
		\addplot[
		color=red,
		mark=square,
		only marks
		]
		coordinates {
			(11,280)(12,286)(13,271)(14,360)(15,478)
		};
		\addplot[
		color=blue,
		mark=o,
		only marks
		]
		coordinates {
			(11,293)(12,316)(13,280)(14,383)(15,466)
		};
		\end{axis}
		\end{scope}
		\end{tikzpicture}
		&
		\begin{tikzpicture}
		\begin{scope}[scale=0.46, transform shape]
		\pgfkeys{{/pgfplots/legend entry/.code=\addlegendentry{#1}}}
		\begin{axis}[
		xlabel={Set},
		legend pos=north east,
		ymajorgrids=true,
		grid style=dashed,
		]
		\addplot[
		color=red,
		mark=square,
		only marks
		]
		coordinates {
			(11,156)(12,191)(13,176)(14,280)(15,382)
		};
		\addplot[
		color=blue,
		mark=o,
		only marks
		]
		coordinates {
			(11,154)(12,171)(13,162)(14,238)(15,328)
		};
		\end{axis}
		\end{scope}
		\end{tikzpicture}
		\\
		(G) & (H) & (I)
		
	\end{tabular}
    \end{adjustbox}
	\caption{Plots (A)-(C) are for sets 1-5 (varying \colorbox{codecolors_inline}{\lstinline{scale}}), (D)-(E) are for sets 6-10 (no cycles, varying \colorbox{codecolors_inline}{\lstinline{n_lat}}), (G)-(I) for sets 11-15 (one cycle, varying \colorbox{codecolors_inline}{\lstinline{n_lat}}).}
	\label{graphs}
\end{figure}
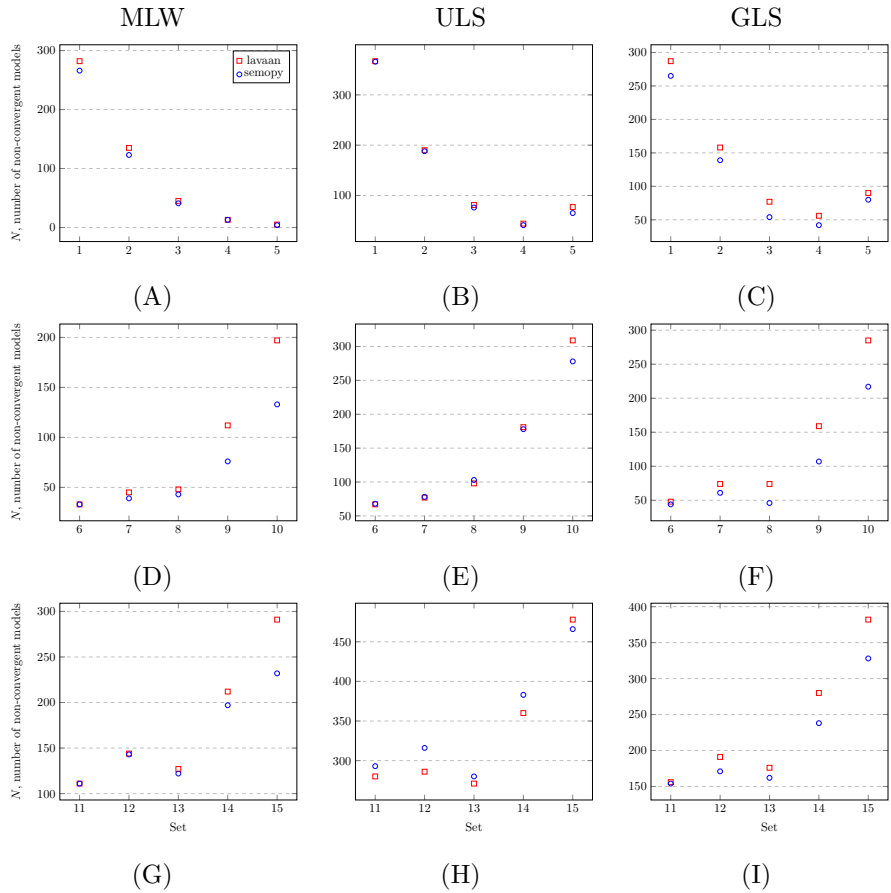

\subsection{Performance}
We've benchmarked both \textbf{semopy} and \textbf{lavaan} on a mixed set of models. We were using a computer with AMD A10-4600M CPU, OS Manjaro 4.14 and \textbf{OpenBLAS} as a backend for \textbf{numpy} and \textbf{scipy}.

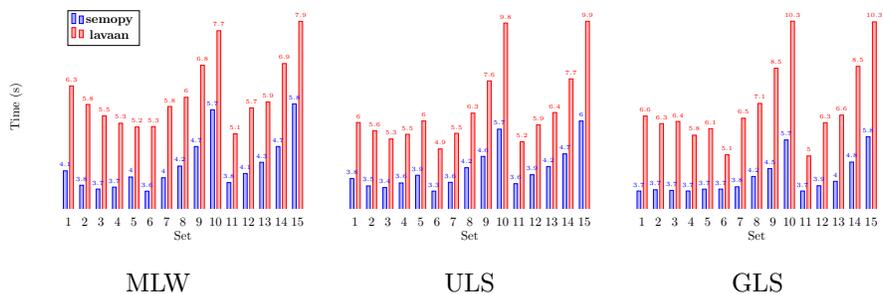
\begin{figure}[H]
	\centering
	\begin{adjustbox}{max width=\textwidth} 
	\begin{tabular}{ccc}
		\begin{tikzpicture}
		\begin{scope}[scale=0.48, transform shape]
		\begin{axis}[
		ybar,
		y axis line style = { opacity = 0 },
		x axis line style = { opacity = 0 },
		xtick style={draw=none},
		ymode=log,
		legend pos=north west,
		enlarge x limits={abs=0.5},
		xtick={1,2,3,4,5,6,7,8,9,10,11,12,13,14,15},
		bar width=3,
		xlabel={Set},
		ylabel={Time (s)},
		ytick=\empty,
		nodes near coords,
		every node near coord/.append style={font=\tiny, /pgf/number format/.cd,fixed,precision=1}
		]
		\addplot coordinates{(1, 62.0) (2, 43.0) (3, 39.0) (4, 41.0) (5, 53.0) (6, 37.0) (7, 52.0) (8, 70.0) (9, 115.0) (10, 293.0) (11, 46.0) (12, 58.0) (13, 77.0) (14, 115.0) (15, 340.0)};
		\addplot coordinates{(1, 537.0) (2, 335.0) (3, 251.0) (4, 208.0) (5, 189.0) (6, 191.0) (7, 318.0) (8, 402.0) (9, 907.0) (10, 2182.0) (11, 159.0) (12, 307.0) (13, 357.0) (14, 948.0) (15, 2771.0)};
		\legend{\textbf{semopy}, \textbf{lavaan}}
		\end{axis}
		\end{scope}
		\end{tikzpicture}
		& 
		\begin{tikzpicture}
		\begin{scope}[scale=0.48, transform shape]
		\begin{axis}[
		ybar,
		y axis line style = { opacity = 0 },
		axis y line = none,
		x axis line style = { opacity = 0 },
		xtick style={draw=none},
		ymode=log,
		enlarge x limits={abs=0.5},
		xtick={1,2,3,4,5,6,7,8,9,10,11,12,13,14,15},
		bar width=3,
		xlabel={Set},
		ylabel={Time (s)},
		ytick=data,
		nodes near coords,
		every node near coord/.append style={font=\tiny, /pgf/number format/.cd,fixed,precision=1}
		]
		\addplot coordinates{(1, 44.0) (2, 33.0) (3, 31.0) (4, 37.0) (5, 50.0) (6, 27.0) (7, 38.0) (8, 67.0) (9, 104.0) (10, 300.0) (11, 36.0) (12, 51.0) (13, 70.0) (14, 115.0) (15, 412.0)};
		\addplot coordinates{(1, 389.0) (2, 283.0) (3, 205.0) (4, 246.0) (5, 413.0) (6, 139.0) (7, 255.0) (8, 560.0) (9, 1967.0) (10, 18599.0) (11, 184.0) (12, 356.0) (13, 573.0) (14, 2124.0) (15, 19969.0)};
		\end{axis}
		\end{scope}
		\end{tikzpicture}
		& 
		\begin{tikzpicture}
		\begin{scope}[scale=0.48, transform shape]
		\begin{axis}[
		ybar,
		y axis line style = { opacity = 0 },
		axis y line = none,
		x axis line style = { opacity = 0 },
		xtick style={draw=none},
		ymode=log,
		enlarge x limits={abs=0.5},
		xtick={1,2,3,4,5,6,7,8,9,10,11,12,13,14,15},
		bar width=3,
		xlabel={Set},
		ylabel={Time (s)},
		ytick=data,
		nodes near coords,
		every node near coord/.append style={font=\tiny, /pgf/number format/.cd,fixed,precision=1}
		]
		\addplot coordinates{(1, 39.0) (2, 41.0) (3, 40.0) (4, 39.0) (5, 42.0) (6, 42.0) (7, 46.0) (8, 69.0) (9, 94.0) (10, 288.0) (11, 39.0) (12, 48.0) (13, 57.0) (14, 121.0) (15, 328.0)};
		\addplot coordinates{(1, 735.0) (2, 547.0) (3, 594.0) (4, 347.0) (5, 449.0) (6, 161.0) (7, 675.0) (8, 1204.0) (9, 4781.0) (10, 30102.0) (11, 154.0) (12, 567.0) (13, 761.0) (14, 5131.0) (15, 29336.0)};
		\end{axis}
		\end{scope}
		\end{tikzpicture}
		\\ 
		MLW & ULS & GLS
	\end{tabular}
    \end{adjustbox}
	\caption{Performance benchmark, logarithmic scale}
	\label{timebench}
\end{figure}


\subsection{Optimisation methods}
All optimisation methods available in the package were also tested against each other.
\begin{table}[H]
	\centering
	\caption{An entry at row $i$ and column $j$ stands for number of models that didn't converge for method $i$ yet did for method $j$. Diagonal entries stand for number of models where methods didn't converge.}
	\label{optimtable}
	\begin{adjustbox}{max width=\textwidth} 
	\begin{tabular}{|c|c|c|c|c|c|c|}
		\hline
		& \textbf{SLSQP} & \textbf{L-BFGS-B} & \textbf{SMSNO} & \textbf{SUMSL} &\textbf{HUMSL} & \textbf{Adam} \\ \hline
		\textbf{SLSQP} & 133       &      7     &     2       &   4     &     11    &    46 \\ \hline
		\textbf{L-BFGS-B}  &  609     &      735  &    152 &   595 &  602   & 449 \\ \hline
		\textbf{SMSNO}     &   603     &     151     &   734   &   583 &     589 &   449 \\ \hline
		\textbf{SUMSL}      &     23          &   12       &   1       & 152      &    9      &  52 \\ \hline
		\textbf{HUSML}   &        23         &    12        &  0       &   2     &   145     &   49 \\ \hline
		\textbf{Adam}   &       255 &          56   &    57   &     242   &     246  &     342 \\ \hline
	\end{tabular}
   \end{adjustbox}
\end{table}
As it can be seen from Table \ref{optimtable}, SLSQP clearly outperforms other methods, however, it has to be noted that there are cases when SLSQP fails to correctly estimate parameters whereas other methods successfully find a solution. Thereof we conclude that other optimisation methods should be given a shot in case of SLSQP's failure. 
%
%

\nocite{*}
\bibliography{bibliography}
\end{document}